\documentclass[conference]{IEEEtran}
\usepackage{cite}
\usepackage{amsmath,amssymb,amsfonts}
\usepackage{algorithmic}
\usepackage{graphicx,dblfloatfix}
\usepackage{textcomp}
\usepackage{xcolor}
 \usepackage[utf8]{inputenc}
 \usepackage{cite}
\usepackage{amsmath,amssymb,amsfonts}
\usepackage{algorithmic}
\usepackage{graphicx}
\usepackage{subcaption}
\usepackage{textcomp}
\usepackage{xcolor}
\usepackage{multirow}
\usepackage{color}
\usepackage{pifont}
\usepackage{float}
\usepackage{algorithm2e}
\usepackage{url}
\usepackage{xcolor}
 \usepackage{multirow}
 \usepackage{dblfloatfix}
 \usepackage{flushend}
\usepackage{placeins}
\PassOptionsToPackage{bookmarks=false}{hyperref}

\begin{document}
\title{SDN-based Self-Configuration for \\ Time-Sensitive IoT Networks}

\author{\IEEEauthorblockN{
 Nuref\c{s}an~Sertba\c{s} B{\"u}lb{\"u}l\ , Do\u{g}analp Ergen\c{c}, Mathias Fischer}
\IEEEauthorblockA{Department of Computer Science,
University of Hamburg, Germany\\
Email:\{sertbas, ergenc, mfischer\}@informatik.uni-hamburg.de
}}

\maketitle
%page numbers
\thispagestyle{plain}
\pagestyle{plain}

\begin{abstract} 
The convergence of IT and OT technologies results in the need for efficient network management solutions for automotive and industrial automation environments. However, configuring real-time Ethernet networks while maintaining the desired QoS is challenging due to the dynamic nature of OT networks and the high configuration parameters. 
This paper introduces an SDN-based self-configuration framework for the fully automated configuration of TSN networks. Unlike standard TSN configuration, we remove end-host-related dependencies and put flows initially on default paths to extract traffic characteristics by monitoring network traffic at edge switches. Communicated to a central SDN controller, these characteristics allow to move the flows to optimal paths while maintaining hard real-time guarantees, for which we also formulate an optimization problem. 
Our simulation results indicate that the proposed self-configuration approach works properly for different network sizes and numbers of end-hosts. Even though it slightly increases the average latency of critical frames, it still provides a certain level of real-time guarantee without any prior knowledge of flows.
\end{abstract}


\section{Introduction} \label{sec1}
The advent of Industry 4.0 and the Industrial Internet of Things (IIoT) enables new manufacturing scenarios that include technologies such as advanced robotics, artificial intelligence, advanced sensors, and cloud computing. In such scenarios, time- and safety-critical messages control physical processes, which means that timely and guaranteed delivery becomes highly significant.
 
 % what is tsn and what is offers
The IEEE 802.1 Working Group proposed Time Sensitive Networking (TSN) standards to empower regular switched Ethernet with real-time (RT) capabilities. As a result, TSN enables the coexistence of critical time-sensitive traffic and traditional Ethernet traffic with various quality of service (QoS) classes such as low-priority and best-effort. Moreover, it offers a wide range of functionalities, such as time synchronization, reliability, scheduling, and network management for RT systems.
 
% existing network management mechanism IEEE 802.1Qcc
In TSN, the management and configuration of a network is described in IEEE 802.1Qcc Stream Reservation Protocol (SRP) Enhancements and Performance Improvements \cite{8514112}. SRP specifies how to schedule a time-sensitive stream allocating required network resources. Moreover, the enhancements in the 802.1Qcc standard defines alternative network configuration and management schemes leveraging SRP.  
% problem statement 
However, all proposed configuration schemes rely on the active involvement of the end-hosts to declare their service characteristics and communication requirements to a centralized or decentralized management component. 
% why their assumption could fail
This approach requires the manual configuration of end-hosts, which can be low-power sensors, cyber-physical systems, or robotics that may not support registration protocols, to request the necessary resources. Besides, for large-scale systems, such a configuration could be burdensome and requires continuous maintenance. 
%This approach relies on a central assumption that we try to overcome with this paper: 
% -- Assumption 1: End-hosts need to be TSN-aware
%End-hosts need to be TSN-aware. End-hosts transmit their traffic requirements to the network to their edge switch before data is transmitted. Then the network will configure switches and reserve %resources according to that requirements. Instead, end-hosts should able to connect the TSN network as if they connect to any regular network. 
%\dci{but why is it important? what is the real problem here?}
However, self-configuration capability of TSN networks is not part of the current standards as it is just assumed that there is some centralized or decentralized network controller. Nevertheless, such a controller is required in large and dynamic TSN scenarios like smart factories and smart cities.
% -- Assumption 2: fixed flow sizes
%Second, TSN assumes that, once resources are assigned, flow resource demands remain fixed. End-hosts would need to explicitely communicate changed requirements.
%It is assumed that all flows are known a-priori and related resources are allocated before the actual communication to guarantee timely delivery. 
%However, this is not valid for future cyber-physical systems such as smart factories and smart cities.
%Such network scenarios may require adding new flows on the fly, which requires a flexible and more dynamic network configuration. 

For this reason, supporting TSN by other networking concepts like SDN seems beneficial \cite{du_herlich_2016,ehrlich2018software,schriegel2018investigation}. Unlike traditional networking, SDN allows to configure communication patterns, and device settings on the fly \cite{gerhard2019software} based on a centralized control plane. It allows to split up flows for a transmission on multiple paths for load-balancing, to use the available bandwidth more efficiently, and to make network-wide configurations such as time-synchronization.
 
%solution
In this study, we introduce an SDN-based self-configuration approach for time-sensitive IoT networks. In our approach, end-hosts do not need to be TSN-aware and they obtain required network resources transparently. We assign edge switches the task to trigger the resource allocation mechanism with automatically-detected traffic patterns. For that, these edge switches learn flow characteristics by monitoring and share them with the controller, which computes the optimal routing and enforces the related flow rules on the fly. Eventually, our contributions are

\begin{itemize}
\item We introduce a self-configuration approach for the automated SDN configuration of TSN networks without introducing a significant delay to the system where only less than {1\%} of time triggered (TT) packets experience further configuration delay.
\item We propose a learning component that detect traffic characteristics and classifies streams as TT and best effort (BE) automatically with  {99.52\%} and {94.84\%} accuracy, respectively. This component eliminates the need for SRP for various scenarios and enables a seamless configuration scheme for the end-hosts.
\item We formulate the time-sensitive optimal routing (TSOR) model as a mixed integer linear problem (MILP) to find the optimal routes including TSN gate configurations for the streams with detected characteristics.
\end{itemize}

%paper organization
The remainder of this paper is structured as follows: Section \ref{sec2} summarizes the related work on current TSN flow registration approaches. Section \ref{sec3} describes the current TSN configuration approaches. In Section \ref{sec4}, we introduce our overall architecture. Then, we evaluate our approach and describe our simulation results in Section \ref{sec5}. Finally, Section \ref{sec6} concludes the paper and summarizes future work.

\section{Related Work} \label{sec2}
In this section, we present the literature survey on the configuration of time-sensitive networks.
%dynamic solutions
Offline scheduling approaches as in \cite{kumar2017end} statically allocate network resources for the given communication patterns, e.g., such as time-triggered traffic. That approach works in certain scenarios, e.g., automotive systems, where the communication flows are already known at design time. However, to meet the high priority QoS requirements of future industrial networks, dynamically routing packets depending on the current state, e.g., switch workloads, requires a dynamic configuration including a dynamic resource allocation.

%general-dynamic conf
For TSN, configuration of the network resources to transfer TT traffic is described in IEEE 802.1Qcc \cite{8514112} on the architectural level. Since it does not provide a concrete specification, the authors of \cite{gerhard2019software} propose a configuration architecture named Software-Defined Flow Reservation (SDFR) that is based on OpenFlow (OF) , which is a protocol to configure forwarding plane in SDN. However, they only describe the essential components as a proof of concept to manage network resources in RT and to register time-sensitive flows while routing and scheduling mechanisms are left as out of scope. In \cite{kobzan2018secure}, a generic concept for secure and time-sensitive communication in industrial networks is described. Similar to \cite{gerhard2019software}, there is not any further evaluation or the details of an implementation. Besides, the configuration of the RT traffic is left as an open issue.

%new flows at runtime
In \cite{danielis2017dynamic}, the authors propose an SDN-based resource allocation mechanism for accommodating new flows at runtime. In some cases accommodating new flow requires the migration of existing flows. For that, they propose two algorithms as direct and indirect flow migration. Even though updating one switch's forwarding table at a time is pointed as a solution for unpredictable network behaviour due to the imperfect synchronization between the switches, the performance is not evaluated. In another study, a flow-specific bandwidth and buffer capacity reservation mechanism is proposed \cite{guck2014achieving}. Global knowledge of the controller is used in routing algorithm to compute an appropriate network configuration. They also simplify the end-hosts by removing synchronization features and employ time-division multiple access (TDMA) mechanism. However, their MILP-based path-finding approach is too complex to deliver results in RT. In \cite{nayak2017incremental}, the authors utilize SDN for configuring time-sensitive flows. They propose a combined routing and scheduling algorithm for incrementally adding or removing time-sensitive flows at runtime. The approach schedules transmission at the edges, which requires only limited schedule updates. While it does not require any configuration on switches, it assumes that hosts have a proper clock synchronization and are involved in the scheduling process.

These studies mostly focus on TT traffic under significant assumptions such as having a-priori information about the traffic and TSN-aware clock synchronized hosts. The authors in \cite{gutierrez2017self,gutierrez2015configuration} propose a concept of a configuration agent including a monitor, an extractor, and a scheduler component to make RT switches self-configurable. However, they consider only TT traffic and left sporadic traffic as future work. Also, they propose an abstract end-to-end architecture and do not evaluate the overall system. 

\begin{figure*}[ht!]
\centerline{\includegraphics[width=0.9\linewidth]{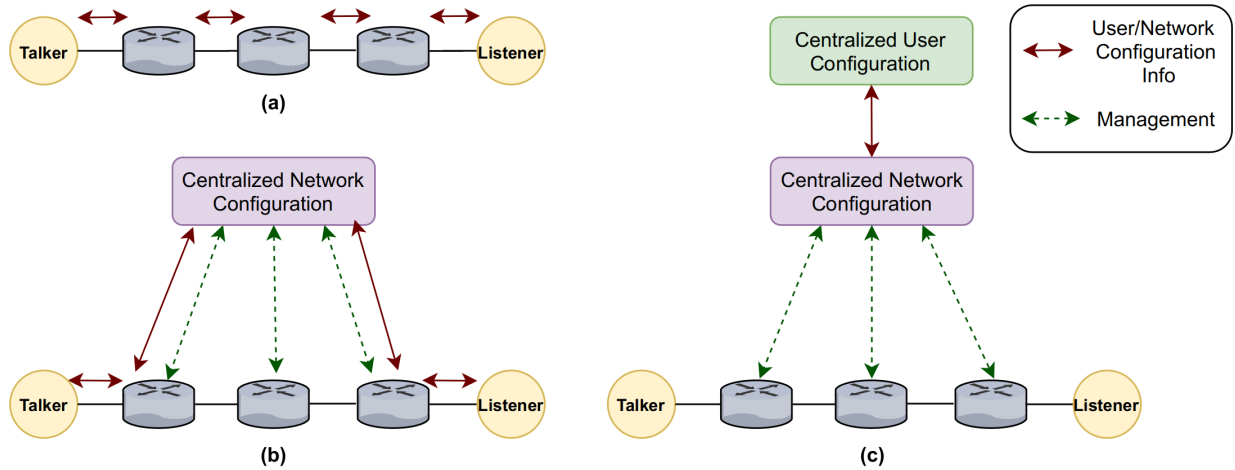}}
\caption{TSN configuration models.}\label{conf}
\end{figure*}

\section{Background on IEEE 802.1Qcc} \label{sec3}

In TSN, the configuration starts at end-hosts named talkers and listeners, which are the source and destination nodes in the TSN context. A talker sends its specific traffic requirements to the edge-switch to request network resources and scheduling. Then, this switch either (i) computes the required resources and scheduling for the related traffic and forwards the request to other switches or (ii) directly forwards the request to a central controller that can configure all the switches on the path towards the listener accordingly. Afterwards, the talker starts sending frames to the network. 

In the rest of this section, the background information on the TSN configuration models is given including the description of the models, user configuration parameters, and the stream reservation protocol.

\subsection{TSN Configuration Models}\label{sec3_1}

There are three models for the configuration of the end-hosts and the network in the current standard. These models describe the logical flow of the configuration information at the architectural level. 

In the \textbf{fully distributed model}, an end-host communicate with the edge switch to declare its traffic requirements and the switch forwards the requirements to the other core switches in the network (See Fig. \ref{conf}-a). Here, switches are not configured by a central entity but in distributed manner with their local knowledge. Such a configuration is not suitable for mechanisms that require collaboration between bridges, e.g., scheduling via time-aware shapers \cite{qbv}.

In the \textbf{centralized network/distributed user model}, user configuration is still distributed and edge switches share the traffic requirements of the end-hosts with a central entity named central network configuration (CNC) instead of propagating them through other bridges (see Fig. \ref{conf}-b). Since some scenarios, such as gate configuration at the switches, require network-wide knowledge and high computational power, CNC offers the better configuration with its global knowledge and possibly higher computational capabilities than forwarding plane elements.

In the \textbf{fully centralized model}, both user and network configurations are centralized by centralized user configuration (CUC) and CNC (see Fig. \ref{conf}-c). End-hosts communicate with the CUC entity to declare their traffic requirements and capabilities. The CUC configures the end-hosts, which is different from previous models involving further interaction with the end-hosts. This might be required for satisfying strict timing requirements by configuring the packet transmission schedules of the end-hosts.

%\dci{Should we relate those models to our approach? Saying that this is different our approach because of that and this...}

\subsection{User/Network Configuration Information}\label{sec3_2}
The TSN user/network configuration information consists of three high-level entities: talker, listener, and status. End-host or the CUC sends a request message that contains the respective talker or listener group. Then, the bridge or CNC sends a reply message that contains the status entity.

The talker entity describes transmission parameters with the following fields:
\begin{itemize}
\item \textit{StreamID}: It is the unique identifier for the Stream and has two elements: The destination MAC address and unique identifier to distinguish between multiple streams originated from the same source MAC address.
\item \textit{StreamRank}: It is used for deciding which streams can be dropped when the network resources reach their limits. 
\item \textit{EndStationInterfaces}: It is a list of physical interfaces that behave as talker/listener.
\item \textit{DataFrameSpecification}: It is used for identifying the frames of a stream based on features like VLAN tag, source, and destination MAC address.
\item \textit{TrafficSpecification}: It specifies how the talker transmits frames for the stream to be used by the controller (or switches) to allocate related resources. It contains fields such as frame size and interarrival time of the frames.
\item \textit{UserToNetworkRequirements}: It specifies stream requirements that need to be satisfied, such as latency and redundancy.
\item \textit{InterfaceCapabilities}: It specifies the network capabilities of all interfaces in the EndStationInterfaces group.
\end{itemize}

Similar to the talker entity, the listener entity also contains \textit{StreamID, EndStationInterfaces, UserToNetworkRequirements} and   \textit{InterfaceCapabilities} features. As a reply to a request from a talker/listener, the following information is provided: 

\begin{itemize}
\item \textit{StreamID}: It identifies the stream for whose status info was provided.
\item \textit{StatusInfo}: It provides the status of the stream configuration.
\item \textit{AccumulatedLatency}: It specifies the worst-case latency for a single frame.
\item \textit{InterfaceConfiguration}: It provides the configuration of interfaces at the talker/listener.
\item \textit{FailedInterfaces}: In case of a failure, it provides a list of interfaces in the failed end-host or switch.
\end{itemize}

\begin{figure*}[t]
\centerline{\includegraphics[width=0.7\linewidth]{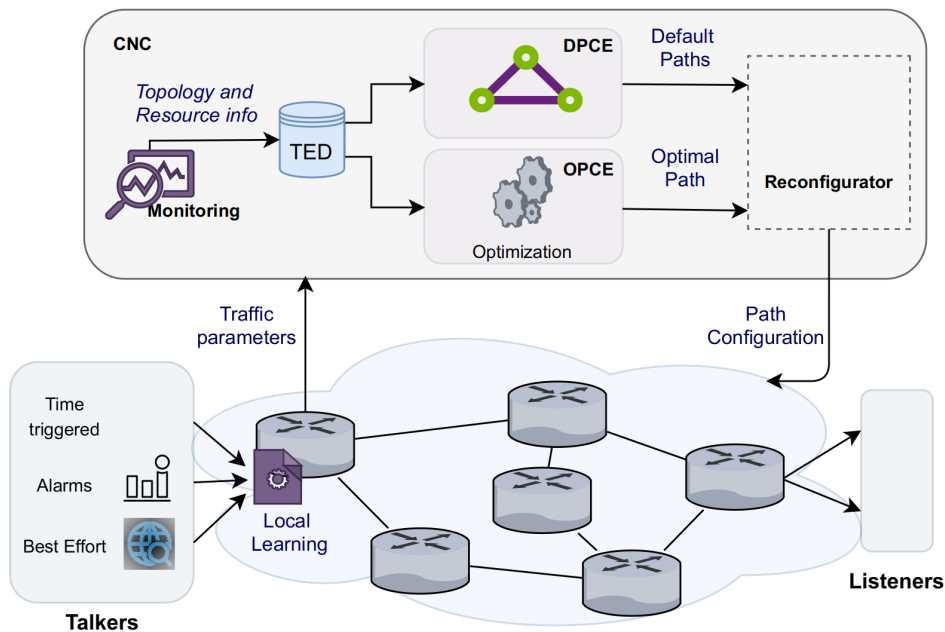}}
\caption{Overall system architecture.}\label{overall}
\end{figure*}

\subsection{Stream Reservation Protocol}\label{sec3_3}

SRP is an extension of the IEEE 802.1Q standard that describes how to manage resource reservations in LANs \cite{8514112}. It defines how to specify and propagate talker registrations through the network with guaranteed QoS. SRP runs at bridges by recording relevant information about the connected end-hosts such as communication latency between a talker and a listener, current stream registrations, etc. The bridges use such information to provide guaranteed QoS for the TSN streams.  
 
SRP can be used in a centralized and a distributed manner as defined in \cite{8514112}. In a distributed model, it only helps to configure a limited number of parameters with the local information in a switch. In the centralized model, SRP can be used to communicate between the talker/listener and CNC. Initially, the talker requests the required bandwidth resources for a stream. As long as there is sufficient bandwidth resources on a selected path from the talker to the listener, that capacity is allocated for the related stream and the switches are configured accordingly. SRP also enables talkers/listeners to later join or leave. However, it requires direct messaging between the end-hosts and the switches. 

As mentioned, SRP requires the active involvement of the end-hosts through that resource reservation process. Here, our goal is to remove such end-host related dependencies. Accordingly, in the next section, we present our TSN self-configuration approach in detail.
  
\section{TSN Self-Configuration Approach} \label{sec4}

In this section, we introduce our SDN-based dynamic self-configuration approach for TSN that we call SC-TSN. Unlike the standard TSN configuration, we remove end host-related dependencies and extract the traffic characteristics in the edge switches. We first describe the overall framework. Then, we explain how we extract traffic characteristics by observation and compute the paths for the TSN flows in detail.

\subsection{SC-TSN Overall System}

For our system design, we follow the distributed user and centralized network configuration model presented in Section \ref{sec3_1}, as shown in Fig. \ref{overall}. In contrast to standard TSN, the end-hosts do not need to communicate to a central entity to announce their traffic requirements. The talker/listeners communicate directly with their edge-switch. The switch extracts traffic requirements and forwards them to the SDN-supported CNC, which is SDN-CNC. The global network view of the SDN-CNC enables highly optimized networks and fast responses to varying demands.

When a new flow arrives at an edge switch, we consider it a low priority BE flow and forward it via the default paths. These paths are computed in the background by the \emph{Default Path Computation Element (DPCE)}. There is also the \emph{Monitoring module} obtains the network topology as well as its resource information, e.g., link utilization, and stores them in the \emph{Traffic Engineering Database (TED)}. We use OF based statistic collection mechanism to keep TED updated.

\begin{figure}[t]
\centerline{\includegraphics[width=.85\linewidth]{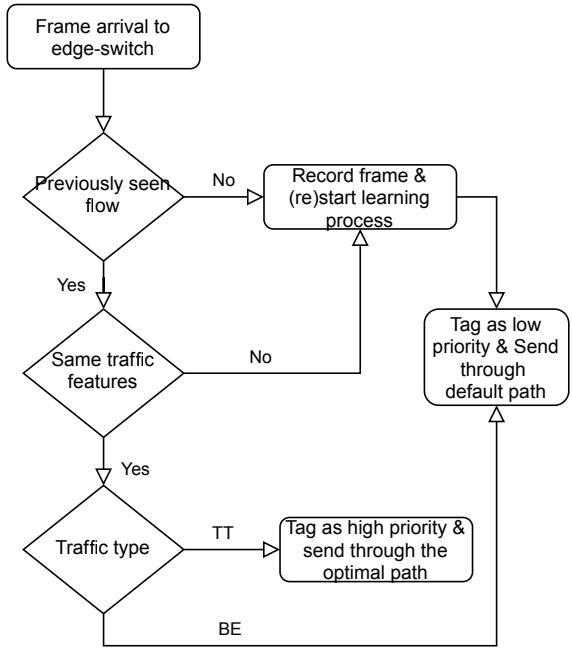}}
\caption{Steps for the flow handling. }\label{fig:flowchart}
\end{figure}

In the meantime, we analyze the received streams at the edge of the network to learn their traffic characteristics to derive their resource and scheduling requirements. For that, we empower edge switches with learning capabilities to extract the traffic patterns such as the frame period $\it{p}$ and the maximum interarrival time of the frames $\it{p_{max}}$.  Those phases are shown in Fig. \ref{fig:flowchart}. All streams are initially tagged as low priority traffic and forwarded via the default paths without resource reservation until we learn their characteristics. If the flow is determined as a TT after a certain time, we tag them as high priority traffic. Then, the \emph{optimal path computation element (OPCE)} computes an optimal path for that flow on the fly, and the flow is migrated to the new path. We also monitor flows to ensure that they have a steady transmission period. In case of that, we calculate deviation from the previously extracted period, restart the learning procedure, and update the configuration.

%\dci{terminology is confusing here. low-priority vs. best-effort, high-priority vs. TT vs. long-living, preconfigured paths vs default paths, streams vs flows}
%\dci{previous two paragraphs have repeating information about how we behave the incoming streams first. but I couldn't untangle them properly...}

%\dci{I think instead of giving many packet fields as bullet points in the previous parts, we should explain these modules.}
% we use stream-->flow
%             bridge-->switch
% 	        preconfigured/previsioned-->default

The SC-TSN aims to eliminate SRP's need as it does not require information about flow priorities beforehand. In other words, we tag flows at the ingress of the network based on the learning module's decision. Besides, SC-TSN also enables SRP-like flow registration procedure via the SDN Northbound API. It could be configured to assign a certain priority level to the particular talker-listener pair. It is possible to configure the network so that either we learn traffic characteristics at the edge or get them via Northbound API. Thus, SC-TSN  is useful for configuration of small to large scale systems where different traffic types such as cyclic/periodic (e.g., signal transmission) or acyclic/sporadic (e.g., event-driven) can coexist. 

\subsection{Learning Traffic Parameters}
As explained in Section \ref{sec3_1}, in the TSN standard, the talker informs the network controller about its traffic requirements before the actual communication starts. That requirement specification includes frame size and interarrival time of the frames, which will be used for the allocation of the required resources. 
Instead, we learn traffic parameters by observing the traffic at the edge switches, enhanced by learning capabilities to analyze incoming frames to extract related parameters. Since we try to learn traffic at the edge, we do not need to consider interference derived from other traffic as in switch to switch links. However, we still need a smart solution here instead of getting average inter-arrival time as a period. 

In the signal processing literature \cite{puech2019fully,gove2018visualizing} analyzing sequences in the frequency domain with Fourier transformation and autocorrelation for periodicity detection is widely used. The Fourier transformation works well for short periods, but may generate many false positives. Thus, the authors of \cite{vlachos2005periodicity} propose to combine Fourier transformation and autocorrelation to detect both short and long periods. In this paper, we use this approach for learning the necessary TSN flow characteristics.

\begin{figure}[t]
\centerline{\includegraphics[width=\linewidth]{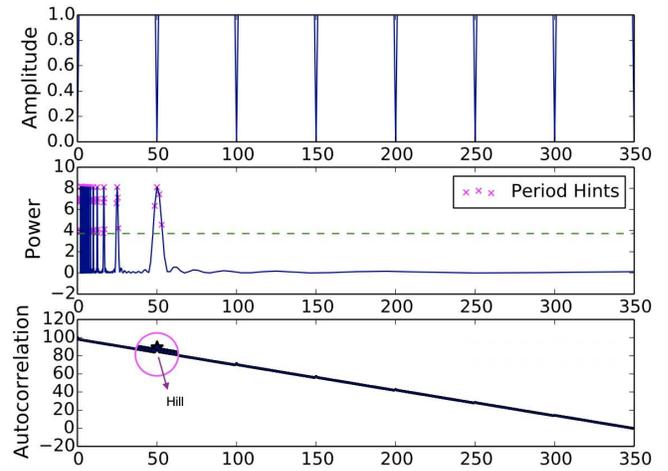}}
\caption{Period extraction steps for example time sequence with period 50.}\label{PeriodMech}
\end{figure}

\begin{figure}[b]
\centerline{\includegraphics[width=\linewidth]{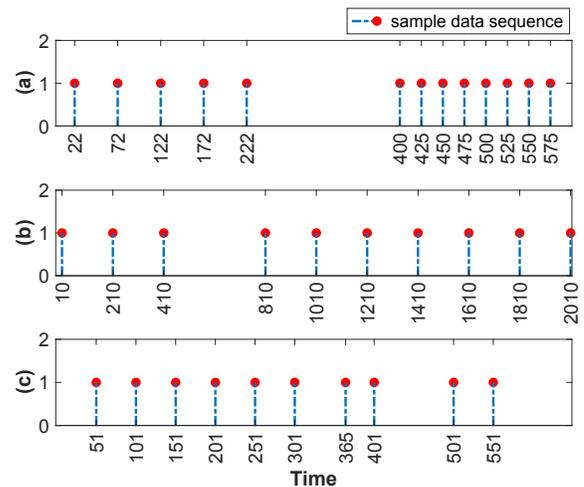}}
\caption{Use case for frequency based period estimator.}\label{usecaseEx}
\end{figure}

We record the arrival time of the frames for each stream and then try to find the period in the frequency domain. For that, we first transform observations to a time sequence $x_{t}=x_{t_{1}},x_{t_{2}},...x_{t_{n}}$ where $x_{t_{k}}=1$ means that a frame arrived at $t_{k}$. Then, we look at the signal's power spectral density by computing the discrete Fourier transform to identify the frequencies that carry most of the energy. In other words, power spectral density analysis can discover the most dominant periods. Then, these periods are validated with the autocorrelation. In that phase, if the candidate period stays at the valley of autocorrelation function, it is interpreted as a false alarm and is discarded. In case that a period stays at the hill of the autocorrelation function, it is considered a reasonable period. Fig. \ref{PeriodMech} shows a working example for a simple time series with period 50. As can be seen from the figure, looking only at the periodogram could be misleading, and validation by autocorrelation helps to get the exact period. In case that the \emph{Learning module} detects periodic behaviour of a time sensitive traffic, it will trigger the \emph{OPCE} for computation by transferring the learned parameters.

%TODO FOR FUTURE
%\dci{why don't we give the actual formula here, omitting the details or theoratical parts? we still have some places in the paper, and none knows about spectral density analysis or autocorrelation?}

%\mci{the following paragraph is actually evaluation. I am not sure if this is the best place to put it}

In order to show different scenarios for frequency based estimator, we generate three different sample data sequences as shown in Fig. \ref{usecaseEx}. In Fig. \ref{usecaseEx}a, we simulate an end-host that initially sends a packet with 50$\mu$s. Then, traffic behavior changes, and it started to send packets with a 25$\mu$s period. There is no packet between t=222$\mu$s and t=400$\mu$s. This could be an example of data for the event-triggered traffic that sends data at different periods depending on events. Here, a frequency-based period estimator could track the exact period while looking only to the mean of the interarrival times gives the period as 62us. Then in Fig. \ref{usecaseEx}b, we simulate a scenario that the packet does not arrive on its period, which should normally be observed at t=610$\mu$s. Another scenario could be a delayed packet as shown in Fig. \ref{usecaseEx}c. The delayed packet was observed at t=365$\mu$s instead of t=351$\mu$s.
In both cases the frequency-based estimator found the exact period (200$\mu$s in (b), 50$\mu$s in (c)), while the mean of the interarrival times method results in the wrong period. From there, we could clearly say that the frequency-based estimator works well for small and large periods, even for changing traffic behaviors.

\subsection{Computing Default and Optimal Paths}
%We propose routing high priority flows via the optimal paths, while forwarding low priority flows on default paths. With this approach, end-hosts are not involved in the configuration anymore. That removes the necessity for a-priori knowledge of the flows in standard TSN. TT traffic would be an excellent example for such a scenario. Since we do not know the traffic characteristics of TT flows beforehand, reserving resources for such flows negatively impacts the bandwidth utilization. Therefore, we first use pre-provisioned and if possible over-dimensioned default paths for such flows. If the respective flow is likely to stay in the network for a certain time, we could migrate that flow to the optimal path. 

This section explains how we use extracted traffic parameters to send packets to the respective paths, either the default path or the optimal path.    

\begin{figure}[b]
\centerline{\includegraphics[width=0.9\linewidth]{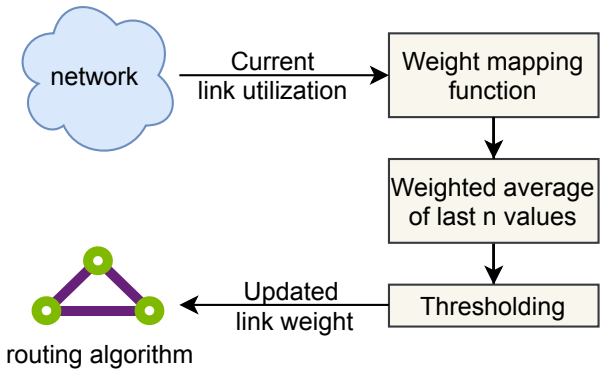}}
\caption{Procedure for updating link weights.}\label{oscillationProblem}
\end{figure}

\subsubsection{Computation of default paths}
To compute default paths for low priority flows, we use a link-utilization based shortest path algorithm. Even tough shortest-path based path computation is widely used, it is not adaptive to the changing traffic since static link weights. Therefore, we changed link weights adaptively based on the current link utilization.

The SDN controller observes the traffic load at the data plane and adjusts the link weights based on the current link utilization. To increase the stability of the forwarding tables and limit the frequent path changes, we follow the same methodology as proposed by \cite{wang2005dynamics}. As shown in Figure \ref{oscillationProblem}, we map current link utilization to the link weights via a linear weight mapping function. Here link weights remain fixed for low utilization values, which keeps the routing overhead low. We compute the weighted average of the last three-link weights, and then we only update the link weights if the change is larger than a threshold, e.g., $20\%$ of its previous value. 

With that, DPCE can dynamically update link weights and computes new paths with the shortest path algorithm. In case that path changes, it will send OF messages to update the flow tables of the switches. Then, these paths are stored at switches to be used for the flows that are tagged as a low priority by the learning module.
%\dci{why don't we write the whole algortihm here, using algorithm environment of latex.}
%this is the least important part of the paper, it should be enough to put few sentences

\subsubsection{Computation of optimal paths}\label{subsec_opt}
The path control and reservation process for TSN is defined in the IEEE 802.1Qca standard. However, it only describes abstract architecture and does not mention any solutions to create such paths. 

\iffalse
\begin{align}
& \sum_{p \in P}^{}{x_{dp}} = 1 & \forall d \in D \label{const:assignment}\\
& \sum_{d \in D}^{}{\sum_{p \in P}^{}{x_{dp} \alpha_{ep} h_d}} \leq c_e & \forall e \in E \label{const:link_capacity}\\
& \sum_{s \in S}^{}{g_{es}} = 1 & \forall e \in E    \label{const:gate_opening}\\
& \sum_{p \in P}^{}{\sum_{e \in E}^{}{x_{dp} \alpha_{ep} \big[l_e^{o} + l_e^{q} (1 - g_{es})\big]}} \leq l_d  \hspace*{-3cm}  &\forall d \in D    \label{const:latency} \\
& g_{es} - \sum_{d \in D}^{}{\sum_{p \in P}^{}{x_{dp} \alpha_{ep} \frac{h_d}{c_e}}} \geq 0 & \forall e \in E, \forall s \in S  \label{const:gate_congestion}\\
& x_{dp} \geq a_{dp} & \forall d \in D, \forall p \in P    \label{const:preassignment}
\end{align}
\fi

Accordingly, in this section, we formulate the time-sensitive optimal routing (TSOR) problem as a mixed integer linear problem (MILP) to migrate high-priority flows to suitable paths.
% what it addresses
Using the model, we find (i) end-to-end paths for given demands under different quality of service (QoS) requirements within limited network resources and (ii) a simple gate configuration per port (iii) minimizing the overall end-to-end communication latency. Note that the gate configuration scheme is not fully considered because the complete scheduling is another complex problem and not in our scope. Here, we calculate how often a particular gate of the respective port of a TSN switch is open instead of the exact configuration of the gate control list (GCL) in the context of time-aware shaper (TAS) \cite{qbv} to satisfy given QoS requirements. 

% variables
Accordingly, we utilize two optimization variables. $x_{dp}$ is a binary variable to decide if demand $d$ is assigned to path $p$. $g_{es}$, is a continuous variable representing the frequency of an open gate on the egress port of link $e$ for the service class $s$ among eight possible classes, including best-effort.  
% explaining g_{es} because it is the tricky part
That is, $g_{es}$ specifies the priority given to service class $s$ on directed link $e$. While $g_{es} = 1$ infers that the gate for $s$ should be open all the time and the entire resource of link $e$ is assigned for that type of demands, $g_{es} \approx 0$ means that any demand of service type $s$ is not active at all on the respective port and thus, the gate is closed. 
% g_es clarified
From this perspective, $g_{es}$ is affected by the total bandwidth required for the demands of service type $s$ as the available resource in $e$ is distributed among those demands according to their service type. It also affects QoS by limiting the forwarding frequency of the packets of such demands. For further information on TAS, service classes and respective gates, and gate control lists, the readers can follow the standard \cite{qbv} and the reference study \cite{Nasrallah2019}.

The constraints and the objective function of TSOR are described below. 
\begin{align}
& \sum_{p \in P}^{}{x_{dp}} = 1 & \forall d \in D \label{const:assignment}
\end{align}
% assignment constraint
Constraint (\ref{const:assignment}) is defined to ensure that each demand $d \in D$ is assigned to exactly one path $p \in P$. Note that we assume here that all flows are non-bifurcated, e.g., not divided into multiple paths.
\begin{align}
& \sum_{d \in D}^{}{\sum_{p \in P}^{}{x_{dp} \alpha_{ep} h_d}} \leq c_e & \forall e \in E \label{const:link_capacity}
\end{align}
% link capacity constraint
Constraint (\ref{const:link_capacity}) is the link capacity constraint and guarantees that each link $e$ has sufficient capacity $c_e$ to handle the total load $h_d$ of all demands $d \in D$ assigned to any path $p$ including $e$, s.t. $\alpha_{ep} = 1$. 
\begin{align}
& \sum_{s \in S}^{}{g_{es}} = 1 & \forall e \in E    \label{const:gate_opening}
\end{align}
% gate configuration constraint
Constraint (\ref{const:gate_opening}) represents the configuration of the gate control list of $e$ for each class of service $s$. Here,  a gate for class $s$ is decided to be open on link $e$ proportional to value of $g_{es}$. As the gates share limited link resources through a network interface (or port), when a set of them are open, others should be closed or open less often.
\begin{align}
& \sum_{p \in P}^{}{\sum_{e \in E}^{}{x_{dp} \alpha_{ep} \big[l_e^{o} + l_e^{q} (1 - g_{es})\big]}} \leq l_d  &\forall d \in D   \label{const:latency}
\end{align}
%latency constraint 
Constraint (\ref{const:latency}) is the latency constraint to ensure that the end-to-end latency on path $p$  is always below the latency requirement of demand $d$, which is $l_d$. Besides, $s$ is the service class of $d$ and the gate configurations $g_{es}$ for that service class through the all link $e$ belongs to path $p$, s.t. $\alpha_{ep} = 1$, impacts the end-to-end latency. 
% justificatio of g_es
Note that while higher values of $g_{es}$ positively impact the latency on link $e$ as it enables the traffic of service type $s$ more often, smaller values of it cause an increased latency due to queueing delay in the respective gate. Thus, a queueing delay factor $l_e^q$ is added proportional to the $1 - g_{es}$, when the gate is closed. Apart from that, a base delay $l_e^o$ representing the port and link characteristics, e.g., packet processing and propagation delay, is considered for each link. While those design parameters, $l_e^q$ and $l_e^o$, can be set according the system and network properties, we use $l_e^q = 0.5$ and $l_e^o = 1.0$ in our simulations.   
\begin{align}
& g_{es} - \sum_{d \in D}^{}{\sum_{p \in P}^{}{x_{dp} \alpha_{ep} \frac{h_d}{c_e}}} \geq 0 & \forall e \in E, \forall s \in S  \label{const:gate_congestion}
\end{align}
% gate congestion constraint
Constraint (\ref{const:gate_congestion}) forces $g_{es}$ to be proportional to the total traffic load of service type $s$ forwarded through the link $e$. Otherwise, it would lead to congestion and unexpected packet drops. 
\begin{align}
& x_{dp} \geq a_{dp} & \forall d \in D, \forall p \in P    \label{const:preassignment}
\end{align}
 % preassignment
Lastly, constraint (\ref{const:preassignment}) fixes the demands that are already assigned to a certain path $p$, i.e., $a_{dp} = 1$ from an existing configuration. $a_{dp}$ is given as input to the problem. 
% small discussion for fixed demands
Note that although keeping the previous demands fixed before allocating a new demand reduces the flexibility of routing, it is important to have a stable configuration scheme especially for the critical and high-priority demands. That is, reconfiguring the network has also a certain cost, e.g., delay for migrating demands, sending control packets to the switches, and can hinder the deterministic communication requirements. The evaluation of that cost might be critical for real deployments but it is out of the scope of this paper.
\begin{align}
\min & \sum_{d \in D}^{}{\sum_{p \in P}^{}{\sum_{e \in E}^{}{x_{dp} \alpha_{ep} \big[l_e^{o} + l_e^{q} (1 - g_{es})\big]}}}  &  \label{const:objective} 
\end{align}
Our objective function (\ref{const:objective}) minimizes the overall latency of the selected paths which is calculated similar to the latency constraint (\ref{const:latency}).

% complexity
Considering the complexity, TSOR has $\mathcal{O}(|D||P| + |E|)$ optimization variables where the number of paths are directly related to the number of links. Note that even though $g_{es}$ depends on the number of service classes, it is, at least in TSN context, defined as eight (including best-effort) and thus we assume that as a constant. In terms of the number of constraints, TSOR is bounded by $\mathcal{O}(|D||P| + |E|)$ constraints with the same assumption on the number of services. 
% linearization
Another important complexity issue is the non-linear constraints and the objective function. It is easily possible to linearize the multiplication of a binary variable $x_{dp}$ and non-binary variable $g_{es}$ using, for instance, McCormick envelopes \cite{Mccormick1976} introducing some additional complexity. Therefore, we take TSOR as a linear problem that makes it more convenient to be solved by the state-of-the-art linear optimization tools.

\begin{figure*}[b]
\centerline{\includegraphics[width=\linewidth]{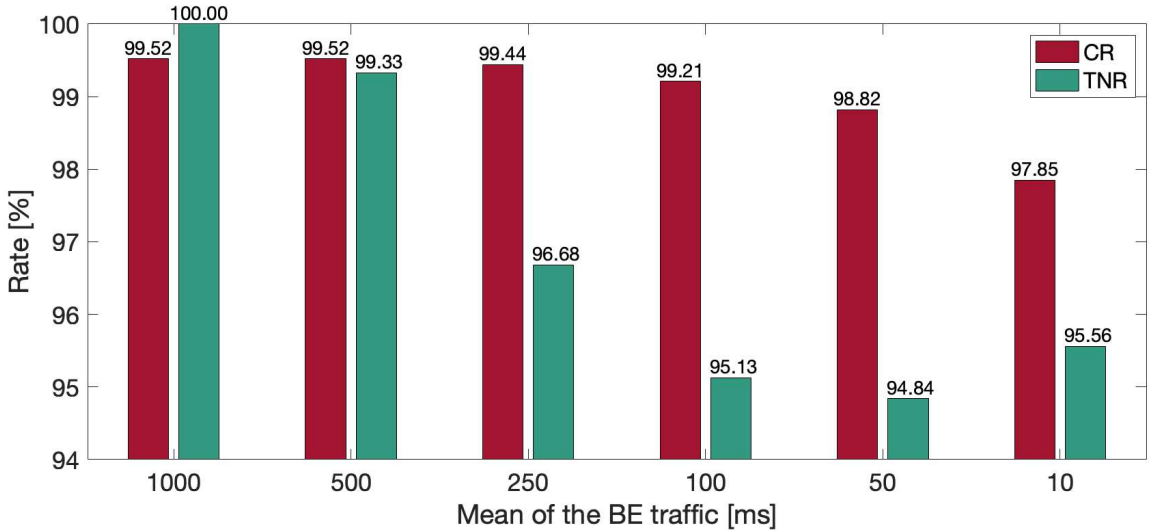}}
\caption{Classification performance of the learning module.}\label{fig:exp0}
\end{figure*}

\begin{table}[b]
\centering
\caption{Topologies used in simulation.}\label{topologies}
\begin{tabular}{|l|c|c|c|}
\hline
\multicolumn{1}{|c|}{\textbf{Metrics $\backslash$ Networks}}   		& \textbf{Getnet} 	& \textbf{Integra} 		& \textbf{Garr201001}  \\ \hline
\textbf{Average node degree}           							& 2.29            		& 2.67         			& 2.52                                                              \\ \hline
\textbf{\# of edge switches}     							& 4               		& 16         		        		& 38                                                                   \\ \hline
\textbf{\# of backbone switches} 						& 3               		& 11          			& 16                                                                   \\ \hline
\textbf{\# of edges between switches}            				& 8              		& 36          			& 68                                                                   \\ \hline
\textbf{\# of hosts per switches}  						& 10              		& 5          				&  2                                                                    \\ \hline
\textbf{Total number of nodes}       							& 47              		& 107          			& 130                                                                 \\ \hline
\end{tabular}
\end{table}

\section{Evaluation} \label{sec5}
In this section, we evaluate SC-TSN and compare it to the SRP-based configuration approach. First, we briefly explain the evaluation setup and evaluation metrics. Then, we evaluate the classification performance of the learning module of our approach at varying traffic load and topology sizes. Finally, we summarize our evaluation results.
	
\subsection{Experimental Setup}
We implemented SC-TSN in the network simulator OMNeT++ v5.5.1 using its INET framework and extending the existing SDN4CoRE framework \cite{hackel2019sdn4core}. SDN4CoRE enables to configure both SDN and TSN capable switches via NETCONF and the OpenFlow protocol. We developed four applications: OPCE, DPCE, Monitoring, and the switch learning module. To find the optimal assignment of flows to the paths, we implemented the optimization problem presented int Section \ref{subsec_opt} in CPLEX 12.7.0. We conducted all experiments in a server with 56-core Intel Xeon 2Ghz CPU and 126GB RAM. 

In our experiments, we used three different real-world network topologies of different sizes from the Topology Zoo dataset: \emph{Getnet}, \emph{Integra}, and \emph{Garr201001}. The characteristics of these topologies are summarized in Table \ref{topologies} \cite{knight2011internet}. We mapped a given topology node to an edge switch with learning capabilities if its node degree is smaller than the average node degree and as backbone switch otherwise. We assumed that end-hosts are connected only to edge switches. 

For all given topologies, we compare our approach with SRP for all given topologies, which is supposed to be the best case.  As explained in Section \ref{sec3_3}, in SRP, everything is given, and all the conditions for an optimal deployment are already there before the actual communication starts. Thus, it is the ultimate competent for SC-TSN.

Different service classes, such as TT traffic and BE traffic, can coexist in the same TSN network. Therefore, we generated mixed traffic scenarios for a comprehensive evaluation. For TT traffic, we uniformly selected talker-listener pairs whose packet sending periods are chosen uniformly between 2-20 ms as stated in \cite{traffic}. We initiated the TT traffic at different time instances and set fixed frame size as 1522 bytes as specified in \cite{hackel2019sdn4core}. the same packet size with TT, i.e., 1522, and exponentially distributed packet interarrival times \cite{alvarez2020comparing,nasrallah2019performance}. We set the same packet generation rate at each BE traffic source and configured them to start transmission at the beginning of the simulation. We also set our simulation time to 100 seconds and set the statistics collection period to 2 seconds for the link weight updates. 

\subsection{Evaluation Metrics}
We use the following metrics to evaluate SC-TSN:
\begin{itemize}
  \item \textbf{End-to-end latency}: The latency of frames until they reach their destination.
  \item \textbf{Number of delayed frames}: The total number of delayed TT frames.
  \item  \textbf{Classification rate (CR)}: The ratio of correctly classified TT and BE frames to the total frames.
  \item \textbf{The true negative rate (TNR)}:  The ratio of correctly classified BE frames.  
\end{itemize}
%\dci{why did we select those metrics, to evaluate what?}
 
\subsection{Results}\label{subsec:res}

In this section, we first present the performance of the learning module at edge switches. Then, we compare SC-TSN with the SRP-based configuration in terms of the end-to-end latency. Finally, we measure the performance of SC-TSN under an increasing number TT streams.

%\dci{we defintely need to clarify what the load is. is the load measure by ms? is that the frequency of the packets? is that the number of streams?}

\subsubsection{Performance of learning traffic parameters} 
In the first experiment, we evaluated the classification performance of our learning module for BE and TT traffic under different BE loads. For that, we kept the same TT flows for each experiment and varied the interarrival time of BE frames and measured CR and TNR. 
Fig. \ref{fig:exp0} shows the accuracy independent of the interarrival time of BE frames. The results indicate that that we can classify both TT and BE frames with the CR between {99.52\%} and {99.85\%} . We also see that the TNR is between and {94.84\%} and {99.52\%} for different rates of BE traffic. 
For light BE load, e.g., when the mean of BE traffic is 1000ms, we can classify almost all BE flows. However, when the interarrival time decreases, our learning approach starts to classify BE frames as TT.  We run these experiments several times directly in the simulation environment because even though we use the same traffic loads, different factors such as queuing delays affect the frames' arrival time. 

To sum up, our results indicate that the BE classification rate does not change significantly with an increasing load.  In case of the misclassification of BE traffic as a TT, we tag that traffic as a high priority and use the optimal paths instead of the default ones. This may decrease the end-to-end latency of BE frames. On the other hand, the misclassification of TT traffic around {$0.5\%$} does not significantly affect end-to-end TT latency because only the first few frames of each TT flow are misclassified. In that case, those frames are tagged as low priority and sent through the preconfigured default paths. 

\begin{table}[t]
\centering
\caption{End-to-end latency of TT frames for varying BE traffic load.}\label{tab:exp1}
\begin{tabular}{|c|c|c|c|c|c|}
\hline
                                                                    & \multicolumn{2}{c|}{\textbf{SRP}}                                                                                                   & \multirow{7}{*}{\textbf{}} & \multicolumn{2}{c|}{\textbf{SC-TSN}}                                                                                                \\ \cline{1-3} \cline{5-6} 
\textbf{\begin{tabular}[c]{@{}c@{}}Mean BE \\ traffic\end{tabular}} & \textbf{\begin{tabular}[c]{@{}c@{}}Mean \\ {[}ms{]}\end{tabular}} & \textbf{\begin{tabular}[c]{@{}c@{}}Max\\ {[}ms{]}\end{tabular}} &                            & \textbf{\begin{tabular}[c]{@{}c@{}}Mean\\ {[}ms{]}\end{tabular}} & \textbf{\begin{tabular}[c]{@{}c@{}}Max\\  {[}ms{]}\end{tabular}} \\ \cline{1-3} \cline{5-6} 
\textbf{10ms}                                                       & 1.31                                                              & 10.52                                                           &                            & 1.35                                                             & 17.30                                                            \\ \cline{1-3} \cline{5-6} 
\textbf{20ms}                                                       & 1.30                                                              & 4.31                                                            &                            & 1.32                                                             & 11.44                                                            \\ \cline{1-3} \cline{5-6} 
\textbf{50ms}                                                       & 1.29                                                              & 2.67                                                            &                            & 1.30                                                             & 8.08                                                             \\ \cline{1-3} \cline{5-6} 
\textbf{100ms}                                                      & 1.29                                                              & 2.48                                                            &                            & 1.30                                                             & 6.24                                                             \\ \cline{1-3} \cline{5-6} 
\textbf{1000ms}                                                     & 1.29                                                              & 2.47                                                            &                            & 1.29                                                             & 5.80                                                             \\ \hline
\end{tabular}
\end{table}

 \begin{figure*}[t!]
    \centering
    \begin{subfigure}[t]{0.5\textwidth}
        \centering
  {\includegraphics[width=0.9\linewidth]{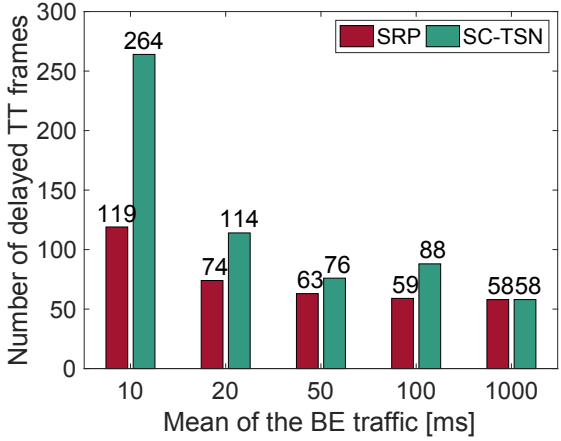}}
  \caption{Number of delayed TT frames.}\label{fig:exp2}
    \end{subfigure}%
    ~ 
    \begin{subfigure}[t]{0.5\textwidth}
        \centering
        {\includegraphics[width=0.9\linewidth]{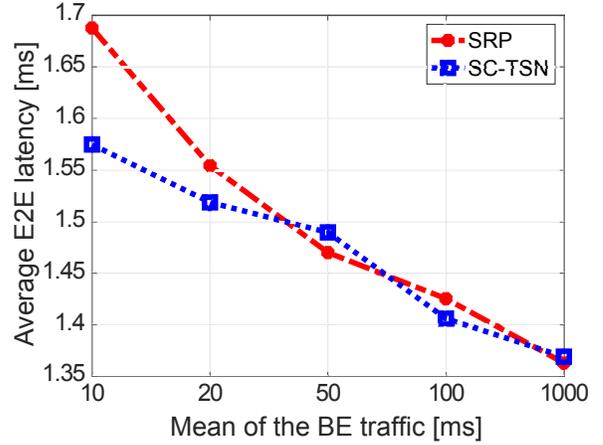}}
  \caption{End-to-end latency of BE frames}\label{fig:exp3}
    \end{subfigure}
\caption{Comparison of SC-TSN with SRP under the varying BE traffic load.}\label{fig:exp1}
\end{figure*}

\subsubsection{Impact of learning on the delivery performance}   

In the next experiments, we compared SC-TSN with the SRP-based configuration that has the traffic requirements of TT flows before the communication. Unlike SRP, we assume that we do not have information about the talker and its flows, and we learn this by observation at edge switches. 
%\dci{assumes to know all TT flow apriori? I know what you meant but reviewer probably wouldn't. I don't change here because I think we can write a more extensive explanation for the reasoning of the selection of SRP as our opponent in one of the previous sections.}
%\dci{we already explained those thing above for a few times, I think here is now not the place to repeat that. let's focus on the results.}

%\dci{why did we select those number 26 and 53}
In the second experiment, we measure how the delay of TT flows is affected by increasing the BE traffic load. We used the Integra topology and determined the number of TT traffic sources as 53 and BE sources as 26 proportional to the number of nodes in the network. Then, we repeated the experiment for different interarrival times ($\mu$) of the BE frames, ranging from 10ms to 1000ms as given in Table \ref{tab:exp1}. We measured the latency and the number of delayed frames. As expected, SRP and SC-TSN overlapped significantly; they have the same average and minimum TT latency values. Since we use priority-based scheduling at the switches, the average latency of the TT frames is not significantly affected by the increasing load of the BE traffic. However, we observe an increase in the maximum latency. Our approach has a higher maximum latency than SRP, because of the learning process. The incoming frames before the extraction of the exact period, are tagged as low priority traffic and send through the default routes. Therefore, they might be significantly delayed. To check this, we measure the number of delayed TT frames, as seen in Fig. \ref{fig:exp2}. In SRP, we observe a lower number of delayed frames. 

As we explained previously, our learning module may classify BE frames as high priority traffic and sends a portion of BE traffic over optimal paths. We see that SC-TSN has lower BE latency than the SRP in between $10$ ms to $50$ms. Even though it seems like the BE classification rate increases in that interval (see Fig. \ref{fig:exp0}), the number of BE frames is also increasing while the number of TT frames remains the same. In other words, the effect of  misclassified BE frames becomes more visible; therefore, we observe lower BE latency in SC-TSN, as shown in Fig. \ref{fig:exp3}.

\begin{figure}[t]
  \centerline{\includegraphics[width=0.9\linewidth]{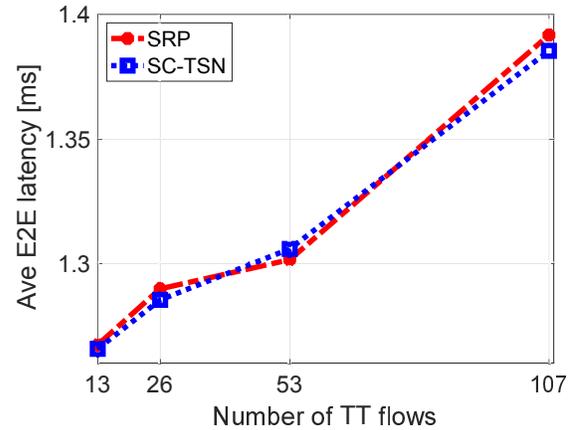}}
\caption{End-to-end latency of TT frames with respect to an increasing number of TT flows.}\label{fig:exp4}
\end{figure}

\begin{figure}[t]
\vspace{-1em}
\centerline{\includegraphics[width=0.9\linewidth]{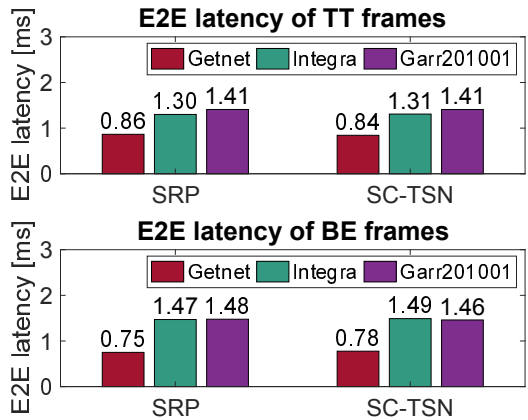}}
\caption{Comparison of SC-TSN with SRP for varied sized topologies.}\label{fig:exp5}
\end{figure}

% \dc{why did we do that? what proportionally?}
In the third experiment, we measure how effectively the OPCE module distributes TT flows and how the increasing number of TT flows affects the latency of TT frames. We set the number of TT flows as ${\{1/8, 1/4, 1/2, 1\}}$ times in proportion to the number of nodes in the network. In the Integra topology, this results in 13, 26, 53, 107 TT flows respectively. We run the experiment under a low BE load by setting $\mu$ to 100 ms, and the number of BE flows to half the number of TT flows. Then, we measure the end-to-end latency of TT flows. Since we use the same OPCE module in both approaches, the results in Fig. \ref{fig:exp4} do not show a significant difference.

In the last experiment, we measure how TT frames are affected by varying sizes of network topologies and the scalability of SC-TSN. We used three different topologies in different sizes as given in Table \ref{topologies}. As in the previous experiments, we set the number of TT flows to half of the number of nodes and the number of BE flows to half of the number of TT flows. Consequently, we have 23 TT and 11 BE sources in Getnet, 53 TT and 26 BE sources in Integra, and 65 TT and 32 BE sources in Garr201001 topologies.  

Fig. \ref{fig:exp5} shows that optimization problem solving time does not significantly affect the end-to-end latency for small topologies such as Getnet. However, for medium- and large-size topologies, Integra and Garr201001 in our setup, we observe that latency increases quickly. One important finding at that experiment is that the latency of TT and BE frames converges in the larger topologies since the solution time of the optimization problem increases with the topology size.

\section{Conclusion} \label{sec6}

Configuration of the time-sensitive networks is a challenging task and requires considerable engineering efforts. Although the alternative configuration schemes have been introduced in IEEE 802.1Qcc standard, self-configuration of TSN is not covered. This paper proposes an SDN-based self-configuration framework for the TSN networks, SC-TSN, in accordance with the plug-n-play nature of Ethernet networks. In that sense, end-hosts do not need to declare their traffic requirements in advance. Instead, the SC-TSN adapts itself for the traffic requirements of the streams with different characteristics and reserve the required resources for routing the data traffic. 

Our experiments indicate that SC-TSN can successfully detect traffic characteristics with over $97.85\%$ classification rate. Moreover, it does achieve results close to the SRP with minimal increase in the end to end latency and below 1\% of the delayed frame rate.

As explained in \cite{qbv}, bounded latency for TT frames can be assured by configuring which 802.1Q priorities are allowed to pass through a particular port at a specific time. Otherwise, end to end latencies are negatively affected by each traversed switches' queuing delays on the multi hop routes. Since we use simple priority-based queuing at the switches instead that kind of time aware configurations, it is difficult to guarantee bounded latency. However, the configuration of gate control lists is possible with the SDN, as shown in \cite{hackel2019sdn4core}. As we consider the gate configuration in our optimization model, TSOR, it is also a valuable future work to extend our whole design, including gate-configuration features.

%jumbo frame
In our current implementation, we considered only the fixed packet sizes but it is also one of the parameters to be used in SRP. Therefore, we should extend SC-TSN in that direction to handle frequently changing packet sizes, including jumbo frames.
%we check the arrival time of every TT frame, whether the end-host is still sending with the same transmission period or not. However, we only consider the traffic whose message sizes are fixed at design time and remain constant for each cycle \cite{traffic}. For instance, we could allocate resources, and then after a certain time, the end-host may reduce the packet sizes. This may results in reserving resources more than we need. Otherwise, we may allocate inadequate capacity for the varied packet sized end host. In both cases, our framework will not work properly. Thus, varying sized frames should also be taken into account in further studies. 

%wrongly 
We also aware that wrongly tagged BE frames use optimal paths and cause waste of resources. Therefore, we would not allocate new resources to TT frames after a certain point. This emphasizes the importance of the learning accuracy. However, it would not be appropriate to deploy complicated mechanisms here due to the run time complexity. There could be other candidates with low computation overhead but high classification accuracy.
%\dci{should we really say that? one can also argue that instead of leaving the paths non-utilized, we should place BE traffic there. it is a trade-off.

\IEEEtriggeratref{18}
\bibliographystyle{IEEEtran}
\bibliography{IEEEabrv,dynamic}
\end{document}